# Chiral Majorana edge state in a quantum anomalous Hall insulator-superconductor structure

Qing Lin He[1]†*, Lei Pan[1]†, Alexander L. Stern[2], Edward Burks[3], Xiaoyu Che[1], Gen Yin[1], Jing Wang[4,5], Biao Lian[5], Quan Zhou[5], Eun Sang Choi[6], Koichi Murata[1], Xufeng Kou[1, 7]*, Tianxiao Nie[1], Qiming Shao[1], Yabin Fan[1], Shou-Cheng Zhang[5], Kai Liu[3], Jing Xia[2], and Kang L. Wang[1]*

[1]Department of Electrical Engineering, University of California, Los Angeles, California 90095, USA.

[2]Department of Physics and Astronomy, University of California, Irvine, California 92697, USA.

[3]Physics Department, University of California, Davis, California 95616, USA.

[4]State Key Laboratory of Surface Physics, Department of Physics, Fudan University, Shanghai 200433, China.

[5]Department of Physics, Stanford University, Stanford, California 94305, USA.

[6]National High Magnetic Field Laboratory, Florida State University, Tallahassee, Florida 32310-3706, USA

[7]School of Information Science and Technology, ShanghaiTech University, Shanghai 200031, China

*Correspondence to: qlhe@ucla.edu; kouxf@shanghaitech.edu.cn; wang@ee.ucla.edu.

†These authors contributed to this work equally.

**After the recognition of the possibility to implement Majorana fermions using the building blocks of solid-state matters, the detection of this peculiar particle has been an intense focus of research. Here we experimentally demonstrate a collection of Majorana fermions living in a one-dimensional transport channel at the boundary of a superconducting quantum anomalous Hall insulator thin film. A series of topological phase changes are controlled by the reversal of the magnetization, where a half-integer quantized conductance plateau ($0.5e^2/h$) is observed as a clear signature of the Majorana phase. This transport signature can be well repeated during many magnetic reversal sweeps, and can be tracked at different temperatures, providing a promising evidence of the chiral Majorana edge modes in the system.**

Viewed as a superconducting analog of the quantum Hall state [1], a chiral topological superconductor (TSC) in two-dimensions can be described by an odd-integer Chern number $\mathcal{N}$ and featured by a full pairing bulk gap with $\mathcal{N}$ gapless chiral Majorana edge modes (CMEMs) [2,3]. The mediator of such edge modes, Majorana fermion, is a putative elementary spin-1/2 quasi-particle with the fascinating character of being its own antiparticle surmised by Ettore Majorana in 1937 [4]. Existences and explorations of Majoranas are intimately steered to the concept of topological phase of quantum matter with earlier representative examples such as the $v = 5/2$ quantum Hall state [5], Moore-Read-type states in the fractional quantum Hall effect [6], vortices in two-dimensional $p+ip$ spinless superconductors [7], strong spin–orbit coupling semiconductor–superconductor heterostructures [8,9], and ferromagnetic atomic chains on a superconductor [10,11], *etc*. The fundamental aspects of Majoranas and its non-Abelian braiding properties can be potentially utilized to implement topological qubits in fault-tolerant quantum computations [12,13], attracting intense attention for years. Starting with the original proposal by Fu and Kane [14,15], numerous schemes to couple topological matters with superconductors [16-27] have been forwarded to construct admixtures accommodating Majoranas, which raised up an upsurge to seek for Majoranas from both theoretical and experimental aspects. Despite Majoranas behave as charge-neutral and zero-energy in a superconductor host, in principle, indication of their existence can be experimentally demonstrated by "zero-bias conductance anomalies" modulated by external electrical/magnetic fields [18-25,28]. So far, such an experimental signature, which has been vastly observed in many quantum systems [18,20,23,25,28], is regarded as the most important evidence originated from zero-dimensional Majorana bound states at the defects/boundaries of a gapped TSC. Although these observations provide promising signatures of Majorana bound states, it is difficult to energetically resolve the contributions from other effects such as Kondo correlations, Andreev bound states, weak anti-localization and reflect-less tunneling, *etc*. [18-20,29-31]. The search for Majorana fermions calls for an experimental observation of Majorana states of higher dimensions, which obey the Majorana equation and hence more faithfully reflect the original idea of Ettore Majorana [4] than bound states. This has never been probed directly or indirectly in any experiments. In this study, we experimentally demonstrate a transport signature of one-dimensional CMEMs in a quantum anomalous Hall insulator (QAHI)-superconductor structure.

TSC can be constructed using the building blocks of topological matters. A typical example is a quantum Hall system with Chern number $\mathcal{C}$ proximately coupled to a superconducting reservoir. However, a strong external magnetic field is usually required to achieve conventional quantum Hall states, which significantly suppresses the superconducting phase [21,27]. Most recently, a series of theoretical results [32-34] suggest that a chiral TSC based on a QAHI might be a promising host of Majoranas since the chiral Hall state can be achieved without the assistance of strong external magnetic fields. In order to break time-reversal symmetry, the single-domain phase of QAHI requires an external field of ~ 0.1T, which can be more than one order of magnitude smaller than the critical field of typical superconducting metals. Modulated by the external field, topological transitions go through CMEMs of both even and odd Chern numbers, leading to the establishment of single CMEMs.

When a superconductor is coupled to a QAHI thin film, *i.e.*, a magnetic topological insulator (TI) thin film, a reversal of the magnetization can induce a series of topological phase transitions. The proposed scheme is demonstrated in Fig. 1a (i)-(vii), where a superconducting region is introduced in the middle of a QAHI channel. The effective Hamiltonian of the QAHI region is written as $\mathcal{H}_0 = \sum_{\mathbf{k}} \psi_{\mathbf{k}}^\dagger H_0(\mathbf{k}) \psi_{\mathbf{k}}$ with $\psi_{\mathbf{k}} = \left( c_{\mathbf{k}\uparrow}^t, c_{\mathbf{k}\downarrow}^t, c_{\mathbf{k}\uparrow}^b, c_{\mathbf{k}\downarrow}^b \right)^T$ and $H_0(\mathbf{k}) = k_y \sigma_x \tilde{\tau}_z - k_x \sigma_y \tilde{\tau}_z + m(k)\tilde{\tau}_x + \lambda \sigma_z$, where $c_{\mathbf{k}\sigma}$ annihilates an electron of momentum **k** and spin $\sigma = \uparrow, \downarrow$; superscripts $t$ and $b$ denote the top and bottom surface states respectively; $\sigma_i$ and $\tilde{\tau}_i$ ($i = x, y, z$) are Pauli matrices for spin and layers while $\lambda$ is the exchange field along the $z$ axis induced by the perpendicular ferromagnetic ordering [33,35]. $m(k) = m_0 + m_1(k_x^2 + k_y^2)$ describes the hybridization between the top and bottom surfaces, which is important for opening a trivial surface gap in the $\mathcal{C} = 0$ state. The Chern number of the system is thus $\mathcal{C} = \lambda/|\lambda|$ for $|\lambda| > |m_0|$ and $\mathcal{C} = 0$ for $|\lambda| < |m_0|$, where $|\mathcal{C}|$ is equal to the number of the chiral edge channels. As a result, by adjusting the external magnetic field, a transition between a normal insulator (NI) with $\mathcal{C} = 0$ (zero plateau, Hall conductance $\sigma_{xy} = 0$) to a QAHI with $\mathcal{C} = \pm 1$ (integer plateau, $\sigma_{xy} = \pm e^2/h$) can be achieved [36,37]. In the middle of the QAHI bar, the proximity to an *s*-wave superconductor drives the QAHI into a superconducting regime, where a finite superconducting pairing amplitude is induced to the surface of the QAHI, and in this case the system can be described by the Bogoliubov-de Gennes (BdG) Hamiltonian $\mathcal{H}_{BdG} = \sum_{\mathbf{k}} \Psi_{\mathbf{k}}^\dagger H_{BdG} \Psi_{\mathbf{k}}/2$, where $\Psi_{\mathbf{k}} = \left[(c_{\mathbf{k}\uparrow}^t, c_{\mathbf{k}\downarrow}^t, c_{\mathbf{k}\uparrow}^b, c_{\mathbf{k}\downarrow}^b), (c_{-\mathbf{k}\uparrow}^{t\dagger}, c_{-\mathbf{k}\downarrow}^{t\dagger}, c_{-\mathbf{k}\uparrow}^{b\dagger}, c_{-\mathbf{k}\downarrow}^{b\dagger})\right]^T$, and

$$H_{BdG} = \begin{pmatrix} H_0(\mathbf{k}) - \mu & \Delta_{\mathbf{k}} \\ \Delta_{\mathbf{k}}^\dagger & -H_0^*(-\mathbf{k}) + \mu \end{pmatrix},$$

$$\Delta_{\mathbf{k}} = \begin{pmatrix} i\Delta_1 \sigma_y & 0 \\ 0 & i\Delta_2 \sigma_y \end{pmatrix}.$$

Here, $\mu$ is the chemical potential and $\Delta_{1,2}$ is the pairing gap function of the top and bottom surface states, respectively [32-34]. In principle, each chiral edge state in a quantum Hall regime is topologically equivalent to two identical copies of CMEMs, such that the total Chern number is even ($\mathcal{N} = 2\mathcal{C}$). The key point to achieve a single CMEM is to induce a topological phase with an odd Chern number such that the two copies of CMEMs can be separated [17]. When structural asymmetry is preserved between the top and the bottom surface states, $\Delta=\Delta_1=\Delta_2$, and the topological transition in the TSC region can only occur between $\mathcal{N} = \pm2$ [Fig. 1a (i), (iv)] and $\mathcal{N} = 0$ (vii), while the QAHI regions experience a NI - QAHI - NI transition due to the interfacial hybridization gap. The topological phase transition for all three regions are synchronized, where the two CMEMs cannot be distinguished from each other. However, due to the broken structural inversion symmetry, the pairing amplitude of the top and the bottom surfaces are different ($\Delta_1 \neq \Delta_2$), such that a phase of $\mathcal{N} = \pm1$ emerges between QAHI$_{\mathcal{C}= \pm1}$ and NI $_{\mathcal{C}= 0}$ ($\Delta = 0$) [32-34].

In order to experimentally identify the CMEM, we fabricate a QAHI-TSC device as illustrated in Fig. 1b. The QAHI bar with dimensions of 2 mm × 1 mm is implemented using a magnetic TI thin film $(Cr_{0.12}Bi_{0.26}Sb_{0.62})_2Te_3$ grown on a GaAs (111)B substrate by molecular beam epitaxy. The Fermi level is within the surface gap without the assistance of electric field tuning [36]. Since the hybridization gap, $m_0$, is important to control the proposed topological phase change, the QAHI film thickness is precisely controlled to be 6 quintuple layers, *i.e.* around 6 nm [36,38]. Across the central part of the QAHI bar, a superconductor bar (8 mm × 0.6 mm) is deposited, which contains a 200 nm layer of Nb protected by a 5 nm layer of Au. A control QAHI Hall bar of the same dimension is also fabricated near the TSC device on the same wafer. The field-dependent total longitudinal conductance ($\sigma_{12}$) is obtained by passing an alternating current ($I_{12}$ in Fig. 1b) through the outer two probes and measuring the potential drop across the inner two probes (1 and 2 in Fig. 1b). In the control QAHI device, transverse and longitudinal resistivity are obtained using the standard Hall bar setup as illustrated on the right side of Fig. 1b. The Hall conductance ($\sigma_{xy}$) of this device is plotted in Fig. 2d, where two intermediate zero plateaus ($\sigma_{xy} \sim 0$) induced by the NI-QAHI transition occurs at the coercive fields $H_C = \pm 150$ mT, indicating the high quality of the QAHI film and the precise control of the thickness and the Fermi level position [36,38].

The single CMEM corresponding to $\mathcal{N} = \pm 1$ can be identified experimentally by a unique transport signature of a half-integer longitudinal conductance plateau ($0.5e^2/h$) during the reversal of the magnetization [32-34]. When the external field is large enough, the device fully reaches the QAHI scheme ($\mathcal{C} = \pm 1$). In the TSC region, both of the two CMEMs exist, forming the phase of $\mathcal{N} = \pm 2$. Since the two CMEMs are topologically equivalent to one QAHI state ($\mathcal{C} = \pm 1$), all incident edge modes can almost perfectly transmit through the device, leading to $\sigma_{12} = e^2/h$, as schematically shown in Fig. 1a (i) and (vii) [compared with (viii) in the QAHI regime]. When the magnetic field reduces, the TSC region experiences the first topological phase change, leading to $\mathcal{N} = \pm 1$, such that one of the pairing CMEM vanishes. Since the QAHI regions are still in the $\mathcal{C} = \pm 1$ phase, the incident QAHI state can only transmit one of the CMEM to the TSC region, while the other CMEM is almost perfectly reflected. This leads to the separation between the two CMEMs in the incident QAHI state [Fig. 1a (ii) and (vi)], such that a half-integer plateau of longitudinal conductance presents. Further reducing the magnetic field drives the QSHI regions to the NI phase ($\mathcal{C} = 0$), where the interfacial hybridization gap shut down the conducting channels ($\sigma_{12} = 0$). Although the TSC region still experiences two more topological transitions ($\mathcal{N} = -1, 0, 1$), no significant transport signatures can be detected in these cases, as shown in Fig. 1a (iii), (iv) and (v). Thus, during every reversal of the magnetization, $\sigma_{12}$ presents a half-integer plateau close to the coercive fields.

The transport channel formed by the CMEM of $\mathcal{N} = \pm 1$ is experimentally demonstrated by the longitudinal conductance signal of the TSC - QAHI device. Figure 2a illustrates the temperature-dependent upper critical field ($\mu_0 H_{C2}^{\perp}$) in the out-of-plane direction of the Nb bar using a standard four-probe method (magenta configuration in Fig. 1b). Here, the zero-resistance temperature (corresponding to the onset of zero resistance) is extracted from the temperature-dependent resistance characteristics under different perpendicular magnetic fields (inset of Fig. 2a). The resulting $\mu_0 H_{C2}^{\perp}$ shows a linear temperature dependence, which follows the standard linearized Ginzburg–Landau theory for two-dimensional superconductors. As a comparison, the temperature-dependent coercive field ($\mu_0 H_C$) of the QAHI derived from the Hall measurements is also plotted in the same figure. As $\mu_0 H_{C2}^{\perp}$ is more than one order of magnitude larger than $\mu_0 H_C$ of the QAHI, the superconducting reservoir is ensured even when the QAHI is driven into a single magnetic domain regime under a large external magnetic field. Down to 350 mK, remarkably, two clear longitudinal conductance ($\sigma_{12}$) plateaus show up at the low-field shoulders of the $\sigma_{12}$ valleys in device I (Fig. 2b). These plateaus occur at ±80 mT, with conductance values of $0.49 e^2/h$. The plateaus at the high-field shoulders cannot be resolved as clear as the low-field ones. They appear as kinks valued at $0.51(\pm 0.05)\ e^2/h$. The pertinent data obtained from other two TSC devices (II and III) exhibit $0.52(\pm 0.06)$ and $0.48(\pm 0.07)\ e^2/h$ conductance plateaus at similar magnetic field ranges. These plateaus are essentially close to the predicted value ($0.5\ h/e^2$) where the transport in the TSC region is dominated by the one-dimensional channel of the CMEM in the $\mathcal{N} = \pm 1$ phase. Despite the fact that Majorana fermions are charge neutral, the two CMEMs on the two edges of the TSC$_{\mathcal{N}=\pm1}$ constitute a coherent charged basis that transmits exactly one half of the incoming charges. As shown in Fig. 2c, the 20 mK result presents distinct plateaus with the quantized value of $0.50(\pm 0.06)\ e^2/h$ at both shoulders of the $\sigma_{12}$ valleys. Outside the half-plateau regimes, when the magnetization is fully saturated, the conductance reaches to a quantized value of $0.98(\pm 0.04)\ e^2/h$, indicating the $\mathcal{C} = \pm 1\ /\ \mathcal{N} = \pm 2$ phase. On the other hand, after the reversal, $\sigma_{12}$ presents two dips corresponding to the NI phase of the QAHI regions as schematically shown in Fig. 1a (iii), (iv) and (v). Ideally, the NI phase should completely shut down the QAHI channels, leading to zero-conductance plateaus. However, residual bulk carriers in the QAHI regions contribute some current due to topologically trivial states, leaving a finite conductance of ~ 0.05 $e^2/h$. Consistent with the phase transitions during the magnetic reversal, the half-integer plateaus (or kinks) locate at every reversal of the magnetization during the hysteresis loop. In these plateaus, the transport in the TSC region is, in principle, purely dominated by a collection of Majorana fermions living in one-dimensional edge modes. This feature is distinct from the "zero-bias conductance anomalies" induced by zero-dimensional Majorana bound states [18,20,25,28], and can act as a hallmark associated with the Majorana edge transport channels in the TSC$_{\mathcal{N}=\pm1}$ phases.

In order to further confirm the transport signature of the CMEMs in the $\mathcal{N} = \pm 1$ phase, measurements at higher temperatures are carried out, and the results are shown in Fig. 3a. As increasing the temperature, the half-integer plateaus gradually narrow down due to the thermally activated bulk carriers. These features completely disappear above 400, 470, and 390 mK for device I (Fig. 3b), II, and III, respectively. In all three devices, the kinks at high-field reversals are always narrower than the low-field plateaus, and they vanish earlier as the increase of the temperature. This discrepancy between low-field and high-field reversals may be attributed to the hindered superconducting phase due to the magnetic field. Although the external magnetic field is lower than the critical field of the Nb bar (as shown in Fig. 2a), the interfacial exchange coupling between the Nb and the magnetic TI may provide a much larger effective field, which suppresses the superconducting phase in the TSC region.

Other than the topological phase change in the TSC region, other effects that might lead to the observed transport signatures are required to be ruled out. Among them, Barkhausen effect [39] is the most possible one. This effect can be characterized by sudden conductance jumps during the magnetization reversal. However, these jumps are caused by rapid changes of magnetic domains during the reversal dynamical processes, which is a random process in both space and time domains. Observables like magneto-resistance and magnetization measurements in Barkhausen effect are irreproducible when cycling the magnetic field sweeps. On the contrary, in the present observation, the half-integer plateaus can be well repeated with clear temperature traces (Fig. 3a). Another important feature is that Barkhausen effect will not be present in the high-field shoulder of the $\sigma_{12}$ valley as the QAHI enters the single domain regime, which is distinct with the narrow plateau/kink observed within this field range in the TSC device. Thus, we claim that the half-plateau conductance is not due to Barkhausen effect. Here we would emphasize that only when backscattering is owing to the TSC$_{\mathcal{N}=\pm1}$ phase does it involve such a half-quantized conductance. Any backscattering processes from the TSC$_{\mathcal{N}=\pm2}$ or the NSC$_{\mathcal{N}=0}$ phase would only involve a considerably small Andreev scattering, which cannot explain the half-integer conductance observed in this experiment.

To summarize, we have demonstrated, for the first time, the observation of conductance quantization originated from the single CMEMs, which provides concrete evidence to the Majoranas-associated edge state transport. These observations are very distinct from the widely explored "zero-bias conductance anomalies" in various superconducting systems and can clearly reveal the physical picture of Majorana scattering. The demonstrated QAHI - superconductor system could act as a prototype of a chiral TSC in two-dimensions, which paves the road to construct more complex layouts to engineer and manipulate Majorana fermions in solid states.

**References:**

1. G. E. Volovik, An analog of the quantum Hall effect in a superfluid $^3$He film. *Sov. Phys. JETP* **69**, 9 (1988).

2. A. P. Schnyder, S. Ryu, A. Furusaki, A. W. W. Ludwig, Classification of topological insulators and superconductors in three spatial dimensions. *Phys. Rev. B* **78**, 195125 (2008).

3. X. L. Qi, T. L. Hughes, S. Raghu, S. C. Zhang, Time-reversal-invariant topological superconductors and superfluids in two and three dimensions. *Phys. Rev. Lett.* **102**, 187001 (2009).

4. E. Majorana, Teoria simmetrica dell'elettrone e del positrone. *Il Nuovo Cimento* **14**, 171-184 (1937).

5. A. Stern, Non-Abelian states of matter. *Nature* **464**, 187-193 (2010).

6. N. Read, D. Green, Paired states of fermions in two dimensions with breaking of parity and time-reversal symmetries and the fractional quantum Hall effect. *Phys.l Rev. B* **61**, 10267-10297 (2000).

7. A. Kitaev, Anyons in an exactly solved model and beyond. *Ann. Phys.* **321**, 2-111 (2006).

8. J. D. Sau, R. M. Lutchyn, S. Tewari, S. Das Sarma, Generic new platform for topological quantum computation using semiconductor heterostructures. *Phys. Rev. Lett.* **104**, 040502 (2010).

9. J. Alicea, Majorana fermions in a tunable semiconductor device. *Phys. Rev. B* **81**, 125318 (2010).

10. S. Nadj-Perge, I. K. Drozdov, B. A. Bernevig, A. Yazdani, Proposal for realizing Majorana fermions in chains of magnetic atoms on a superconductor. *Phys. Rev. B* **88**, 020407 (2013).

11. B. Braunecker, P. Simon, Interplay between classical magnetic moments and superconductivity in quantum one-dimensional conductors: toward a self-sustained topological Majorana phase. *Phys. Rev. Lett.* **111**, 147202 (2013).

12. A. Y. Kitaev, Fault-tolerant quantum computation by anyons. *Ann. Phys.* **303**, 2-30 (2003).

13. B. I. Halperin *et al.*, Adiabatic manipulations of Majorana fermions in a three-dimensional network of quantum wires. *Phys. Rev. B* **85**, 144501 (2012).

14. L. Fu, C. L. Kane, Superconducting proximity effect and majorana fermions at the surface of a topological insulator. *Phys. Rev. Lett.* **100**, 096407 (2008).

15. L. Fu, C. L. Kane, Probing neutral Majorana fermion edge modes with charge transport. *Phys. Rev. Lett.* **102**, 216403 (2009).

16. X.-L. Qi, S.-C. Zhang, Topological insulators and superconductors. *Rev. Mod. Phys.* **83**, 1057-1110 (2011).

17. X.-L. Qi, T. L. Hughes, S.-C. Zhang, Chiral topological superconductor from the quantum Hall state. *Phys. Rev. B* **82**, 184516 (2010).

18. V. Mourik *et al.*, Signatures of Majorana fermions in hybrid superconductor-semiconductor nanowire devices. *Science* **336**, 1003-1007 (2012).

19. S. Nadj-Perge *et al.*, Observation of Majorana fermions in ferromagnetic atomic chains on a superconductor. *Science* **346**, 602-607 (2014).

20. A. Das *et al.*, Zero-bias peaks and splitting in an Al–InAs nanowire topological superconductor as a signature of Majorana fermions. *Nat. Phys.* **8**, 887-895 (2012).

21. A. R. Akhmerov, J. Nilsson, C. W. Beenakker, Electrically detected interferometry of Majorana fermions in a topological insulator. *Phys. Rev. Lett.* **102**, 216404 (2009).

22. K. T. Law, P. A. Lee, T. K. Ng, Majorana fermion induced resonant Andreev reflection. *Phys. Rev. Lett.* **103**, 237001 (2009).

23. J. Liu, A. C. Potter, K. T. Law, P. A. Lee, Zero-bias peaks in the tunneling conductance of spin-orbit-coupled superconducting wires with and without Majorana end-states. *Phys. Rev. Lett.* **109**, 267002 (2012).

24. J. J. He *et al.*, Correlated spin currents generated by resonant-crossed Andreev reflections in topological superconductors. *Nat. Commun.* **5**, 3232 (2014).

25. J. P. Xu *et al.*, Experimental detection of a Majorana mode in the core of a magnetic vortex inside a topological insulator-superconductor $Bi_2Te_3$/$NbSe_2$ heterostructure. *Phys. Rev. Lett.* **114**, 017001 (2015).

26. M. X. Wang *et al.*, The coexistence of superconductivity and topological order in the $Bi_2Se_3$ thin films. *Science* **336**, 52-55 (2012).

27. Y. Tanaka, T. Yokoyama, N. Nagaosa, Manipulation of the Majorana fermion, Andreev reflection, and Josephson current on topological insulators. *Phys. Rev. Lett.* **103**, 107002 (2009).

28. L. Maier *et al.*, Induced superconductivity in the three-dimensional topological insulator HgTe. *Phys. Rev. Lett.* **109**, 186806 (2012).

29. P. A. Lee, Seeking out Majorana under the microscope. *Science* **346**, 545-546 (2014).

30. R. Žitko, J. S. Lim, R. López, R. Aguado, Shiba states and zero-bias anomalies in the hybrid normal-superconductor Anderson model. *Phys. Rev. B* **91**, 045441 (2015).

31. J. D. Sau, P. M. Brydon, Bound states of a ferromagnetic wire in a superconductor. *Phys. Rev. Lett.* **115**, 127003 (2015).

32. S. B. Chung, X.-L. Qi, J. Maciejko, S.-C. Zhang, Conductance and noise signatures of Majorana backscattering. *Phy. Rev. B* **83**, 100512 (2011).

33. J. Wang, Q. Zhou, B. Lian, S.-C. Zhang, Chiral topological superconductor and half-integer conductance plateau from quantum anomalous Hall plateau transition. *Physical Review B* **92**, 064520 (2015).

34. B. Lian, J. Wang, S.-C. Zhang, Edge-state-induced Andreev oscillation in quantum anomalous Hall insulator-superconductor junctions. *Phys. Rev. B* **93**, 161401(R) (2016).

35. R. Yu *et al.*, Quantized anomalous Hall effect in magnetic topological insulators. *Science* **329**, 61-64 (2010).

36. X. Kou *et al.*, Metal-to-insulator switching in quantum anomalous Hall states. *Nat. Commun.* **6**, 8474 (2015).

37. Y. Feng *et al.*, Observation of the zero Hall plateau in a quantum anomalous Hall insulator. *Phys. Rev. Lett.* **115**, 126801 (2015).

38. J. Wang, B. Lian, S.-C. Zhang, Universal scaling of the quantum anomalous Hall plateau transition. *Phys. Rev. B* **89**, 085106 (2014).

39. H. Barkhausen, *Physikalische Zeitschrift* **20**, 401 (1919).

**Acknowledgments:**

This work was supported as part of the SHINES Center, an Energy Frontier Research Center (EFRC) funded by the U.S. Department of Energy (DOE), Office of Science, Basic Energy Sciences under Award # S000686 S000686, and the National Science Foundation (DMR-1411085, ECCS-1232275, and DMR-1543582). We are grateful to the support from the ARO program under contract 15-1-10561. We also acknowledge the support from the FAME Center, one of six centers of STARnet, a Semiconductor Research Corporation (SRC) program sponsored by MARCO and DARPA. J.W., B.L., Q.Z., and S.C.Z. acknowledge the support from the U.S. DOE, Office of Basic Energy Sciences, Division of Materials Sciences and Engineering, under contract No. DE-AC02-76SF00515. J. W. also acknowledges the support from the National Thousand-Young-Talents Program.

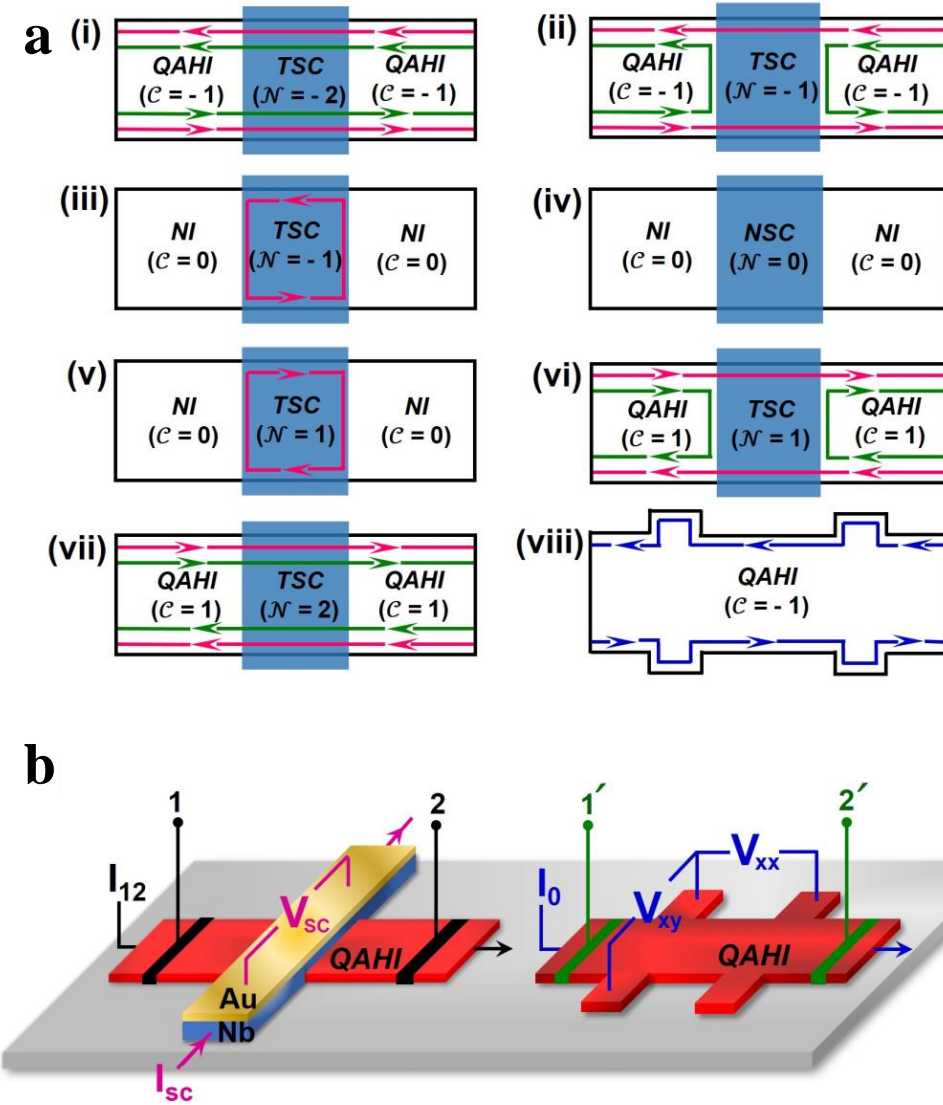

**Fig. 1. Chiral Majorana edge modes (CMEMs) in the quantum anomalous Hall insulator-superconductor structure.** (**a**) The edge transport configurations of the topological superconductor (TSC) device as shown in the left of (**b**) evolved under an external magnetic field. Pink and green arrows represent the chiral Majorana edge modes. There is no backscattering for TSC$_{\mathcal{N}=\pm2}$ and Majoranana backscattering of TSC$_{\mathcal{N}=\pm1}$. The last Hall bar demonstrates the example of the $\mathcal{C}=-1$ chiral edge transport in a QAHI. (**b**) Schematics of a (TSC) device consisting of a quantum anomalous Hall insulator (QAHI) (Cr$_{0.12}$Bi$_{0.26}$Sb$_{0.62}$)$_2$Te$_3$ thin film (6 nm thick) and a superconductor Nb bar. A QAHI Hall bar was also fabricated on the same wafer as a reference. The four-terminal longitudinal conductance ($\sigma_{12}$) of the TSC device was obtained by passing a current ($I_{12}$) and measuring the potential drop across points 1 and 2. To characterized the upper critical field ($\mu_0 H_{C2}^{\perp}$) of the Nb bar, its temperature-dependent resistance was measured using an independent four-probe method by passing a current $I_{SC}$ and measuring the voltage $V_{SC}$ under different perpendicular magnetic field (results also see the inset of Fig. 2a). On the right side, to obtain the Hall conductance ($\sigma_{xy}$) of the QAHI device, longitudinal ($V_{xx}$) and transverse ($V_{xy}$) voltages were measured when passing a current $I_0$. The potential drop across the entire Hall bar ($V_{12}'$) was also independently measured in order to calculate its total longitudinal conductance ($\sigma_{12}'$).

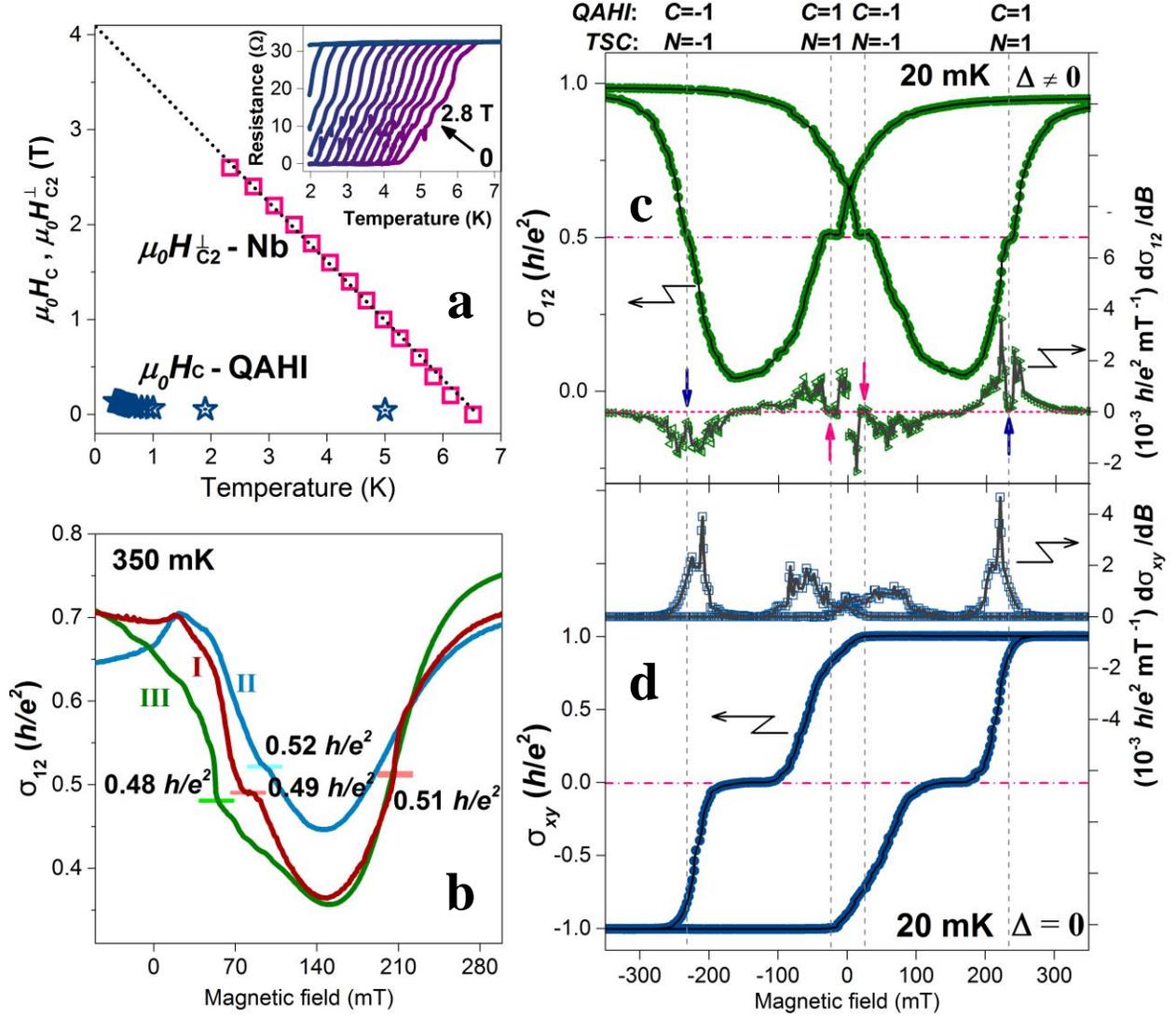

**Fig. 2. Half-integer longitudinal conductance as a signature of single chiral Majorana edge modes.** (**a**) Temperature-dependent perpendicular upper critical field ($\mu_0 H_{C2}^{\perp}$, pink) of the Nb bar and coercive field ($\mu_0 H_C$, cyan) of the QAHI, in which the former was derived from the onset temperature of the zero resistance state under various perpendicular magnetic fields (inset: temperature-dependent resistance curves under magnetic field from zero to 2.8 T in 200-mT steps) while the latter was extracted from the magnetic field dependences of Hall resistance at different temperatures. Standard linearized Ginzburg–Landau theory for superconducting film was used for the fitting of $\mu_0 H_{C2}^{\perp}$ (dotted line). The giant contrast between $\mu_0 H_{C2}^{\perp}$ and $\mu_0 H_C$ ensures the superconducting proximity effect from Nb to QAHI. (**b**) The longitudinal four-terminal conductance ($\sigma_{12}$) of three representative TSC devices as functions of perpendicular magnetic fields at 350 mK. The three devices all show conductance plateaus at the low-field shoulders of the $\sigma_{12}$ valleys with values close to the predicted $0.5e^2/h$, supporting the existence of single CMEMs. Only device I exhibits a sudden increase of $\sigma_{12}$ at the high-field shoulder of the $\sigma_{12}$ valley. (**c**) When superconductivity is induced on the top surface of the QHAI, $\sigma_{12}$ shows additional half-integer plateaus (~ $0.5e^2/h$) between the transitions of the QAHI$_{C=\pm1}$ and normal insulator. This is further illustrated by the derivative of $\sigma_{12}$ with respect to magnetic field (marked by dash lines and arrows), highly consistent with the theoretical prediction of the single CMEMs. (**d**) Without superconducting proximity effect, the Hall conductance ($\sigma_{xy}$) of the QAHI demonstrates the evolution between QAHI$_{C=\pm1}$ and normal insulator, clearly illustrating the surface-hybridization-induced zero plateaus at the coercive fields during magnetization reversals.

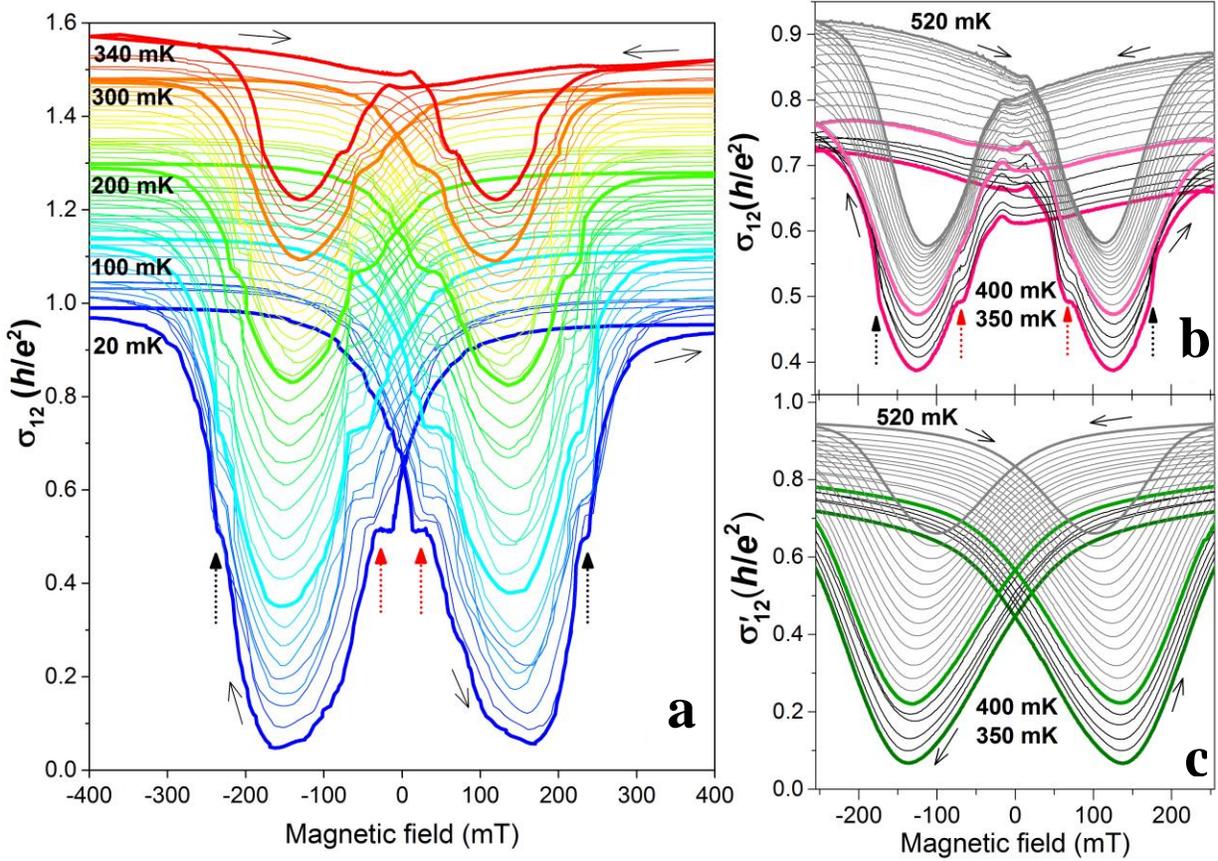

**Fig. 3. Temperature evolution of the total longitudinal conductance in a TSC device.** Four-terminal magneto-conductances in (**a**) and (**b**): a TSC device I ($\sigma_{12}$), and (**c**): a QAHI device ($\sigma_{12}'$), as functions of perpendicular magnetic fields at different temperatures. Traces are offset for clarity, except for the lowest traces at 20 mK in (**a**), 350 mK in (**b**), and 350 mK in (**c**) with 10-mK steps. To clearly demonstrate the temperature dependence, traces are bolded accompanied with corresponding temperatures on the left. The half-integer plateaus gradually narrow down due to the thermally activated bulk carriers and completely disappear above 400 mK for device I as shown in (**b**) [compared with (**c**)]. Black and red dashed arrows indicate the half plateaus positions and solid black arrows for the field-scan tracking.